\documentstyle[preprint,aps,prl]{revtex} 

\begin{document}                       
\draft                                   
\title{Quasi-static cracks and minimal energy surfaces}

\author{V.I. R\"ais\"anen$^{1,2\thanks{Current address}}$, 
E.T. Sepp\"al\"a$^{1}$,
M.J. Alava$^{1,3-4}$, 
and P.M. Duxbury$^4$}

\tighten

\address{
$^1$  Laboratory of Physics,
Helsinki University of Technology,
P.O. Box 1100,
HUT 02015,
Finland }
\address{
$^2$ ICA1,
University of Stuttgart,
Pfaffenwaldring 27,
D-70569 Stuttgart,
Germany
}

\address{
$^3$ NORDITA, Blegdamsvej 17, DK-2100 Copenhagen, Denmark
}

\address{
$^4$  Dept. of Physics/ 
Astronomy and Center for Fundamental Materials Research,\\
 Michigan State Univ., E. Lansing, MI 48824-1116, U.S.A
}

 \date{\today}

\maketitle

\begin{abstract}

We compare the roughness of minimal energy(ME) surfaces and
scalar ``quasi-static'' fracture surfaces(SQF).   
Two dimensional ME and SQF surfaces have {\it the same roughness scaling},
$w\sim L^{\zeta}$ ($L$ is system size) with $\zeta = 2/3$.
 The 3-d ME and SQF results at strong disorder 
 are consistent with the random-bond Ising exponent $\zeta (d\ge 3) 
 \approx  0.21(5-d)$ (d is bulk dimension).  However
 3-d SQF surfaces  are {\it rougher} than ME ones due to 
 a larger prefactor. ME surfaces undergo a ``weakly rough'' 
 to ``algebraically rough'' transition in 3-d, suggesting
 a similar behavior in fracture. 
 \end{abstract}
\pacs{62.20.Mk, 03.40.Dz, 46.30.Nz, 81.40.Np}
  Fracture\cite{n1} continues to attract the attention of the materials
 theory community, with the full spectrum of theoretical
 tools currently being applied to its 
 analysis\cite{n1,n2,n3,n4,n5,n6}.
 Cracks are usually self-affine and their roughness can 
 be characterised by a roughness exponent ($\zeta$), which may
 take on a few distinct values\cite{n8,n9,n10}(here we calculate the
 ``out-of-plane roughness'' of fracture surfaces).
 However, a debate continues about
 whether or not fracture surfaces are ever 
 generated by ``quasi-static'' processes
 \cite{n2,n8,n9,n11}.
 A ``quasi-static'' fracture process is one in which the stress
 field is always close to equilibrium.  For this to be true,
 damage must evolve much more slowly than the time required
 for the stress field to equilibrate.  Slow crack growth and high cycle
 fatigue are expected to be in this limit.  As well as their fundamental
 interest the latter processes are of enormous industrial importance.
 Bouchaud et al.\cite{n9} argue that at short length scales (as probed by e.g.
 Scanning Tunneling Microscopy) quasistatic processes dominate,
 while at longer length scales dynamical processes are of primary
 importance.   Here we present extensive numerical results
 on the topology of quasi-static fracture surfaces in random fuse
 networks.  We also compare these fracture surfaces with minimal
 energy surfaces in the {\it same networks}.  Using fast
 optimization methods, we are able to simulate the latter
 interfaces for large system sizes.  
 
 Surprisingly\cite{n12}, the roughness exponent of
 minimal energy(ME)\cite{n13} surfaces
  and scalar quasi-static brittle fracture surfaces(SQF) have been
  shown to be close
  in two dimensions.  This is surprising because a minimal
  energy surface is the surface of minimum energy in, for example,
  an Ising model with random bonds(see below), which seems to have
  little to do with fracture. Nevertheless there is some experimental
  evidence that this holds in two dimensions \cite{paperi}.
 We  present precise numerical confirmation of the
 equivalence of ME and SQF roughness exponents in two
 dimensions.
   We analyse networks with either continuous or discrete disorder  and
 find the same size dependence in all cases,  in
 contrast to previous suggestions  
 that discrete disorder is special\cite{n13}.
 
  In three dimensions, our calculations
 are for  ME and SQF surfaces of the cubic lattice in the
 \{100\} direction. 
  We choose the low energy,  \{100\} direction 
 as it is more typical of the orientation of fracture surfaces.
 It has been recently claimed that vector quasistatic fracture
(VQF) surfaces are logarithmically rough at
 {\it weak disorder}\cite{n11}.  In the same paper it was
 stated that SQF are algrebraically rough in the same disorder regime. 
  Since this result is at
   odds with the experimental data\cite{n9}, which are naturally for the
   vector case, it was further suggested that 
   quasistatic fracture is not relevant in real experiments.
      The latter statement is difficult
   to believe in, for example, the case of high cycle fatigue, in
   which the time to failure is days to years.  One possible
   reason for this dichotomy is that
   the calculations of ref. \cite{n11}
   may apply at weak disorder and that there
   may be a transition to an algrebraically rough phase as disorder
   increases.  To illustrate that this may occur even in the
  scalar case, we present extensive data confirming our
  claim\cite{n14} that {\it at weak disorder}  ME surfaces
 in the \{100\} orientation are quite flat (probably logarithmically
 rough), while at strong disorder they become
 algebraically rough. 
 
  The simplest realization of a ME surface 
 is a domain wall in a random-bond Ising model with Hamiltonian
 $H = -\sum J_{ij}S_iS_j$,
 where $S_i = \pm 1$ and $J_{ij}$ are random but non-negative.  A domain
 wall is imposed by fixing one face of a square or cubic lattice to
 be positive and the opposite face to be negative. 
 It was only recently 
 realized that the problem of finding the minimal energy 
 surface(domain wall) is equivalent to an important  and well-known problem in
 graph theory(the min-cut/max-flow problem)\cite{n15,n14}.  In the flow
 problem each bond of the lattice has a ``flow capacity'' 
 ($c_{ij}=2J_{ij}$) and
 once the flow capacity of a bond has been
 reached, the excess flow must be shunted along alternative 
 paths(bonds). The domain wall energy is equal to the maximum flow
 that can be pushed through the network without exceeding any
 of the capacities.  Fast, exact algorithms
 are available for the flow problem and we have developed and
 applied these algorithms  previously\cite{n14}. 
  Here we extend these calculations,
 and compare them with results for SQF surfaces.
 
 The fuse network is an electrical network (e.g. a simple cubic lattice)
 in which each bond is a resistor which fails when more than a threshold
 current $i_{ij}$ amp passes through it.  We use the standard {\it hottest
 bond} algorithm for calculating fracture surfaces in the fuse network.
 These electrical networks usually capture many of the essential features of
 failure problems and have become a paradigm in the area\cite{n2}.
 
 To quantitatively compare fracture and minimal energy surfaces, 
 we set the critical current for the failure of a fuse ($i_{ij}$) 
 to the flow capacity used in the minimal surface calculation 
 (i.e. $c_{ij}=i_{ij}$). The simple physical difference between the two cases
   is that in the fuse network, a
 bond carries no current once $i_{ij}$ is exceeded (it breaks),
 while in the minimal surface case a bond continues to carry its
 threshold capacity 
 $i_{ij}=c_{ij}$ even after its threshold is exceeded 
 (it ``yields'' but does not break).  The latter case
 corresponds to a ``perfectly plastic'' response.   
  In all cases, we have done calculations for two types of
   disorder: random dilution, with probability $p$ of a bond being
   present with $2J_{ij} = c_{ij} = i_{ij} = 1$ and; for a uniform distribution 
   with $2J_{ij} = c_{ij} = i_{ij}$ drawn from a uniform distribution 
   with mean $1$ and extending from $1-R$ to $1+R$.  
   
 Results for two dimensions are
 presented in Figs. 1 and 2. 
 In Figs. 1a(dilution disorder) and 2a(continuous disorder), 
 we present data for the size dependence of the
 interface roughness or ``width'' $w \sim c(p) L^{\zeta}$,
  $w^2 = <h^2(x) - <h(x)>>^2 $ ($h(x)$ describes
  the interface position for a given disorder configuration).
 The data in these figures demonstrates that 
 the roughness exponent of minimal energy surfaces and
 scalar fracture surfaces are asymptotically the same in 
 two dimensions. Thus as suggested previously\cite{n12}
 the roughness exponent takes the value
 $\zeta = 2/3$ (dotted line in Figs. 1a and 2a), 
 which is exact for ME's \cite{Kardar}. 
 In Fig. 1b we show that ME and SQF  surfaces
 are rough for arbitrarily weak dilution disorder in two dimensions. 
 We also calculated the roughness of 
 cracks  grown from an initial notch and find results
 consistent with Figs. 1a and 2a. For a given realisation, 
 ME's and SQF surfaces may still be quite different 
 and  have different roughness (see Fig. 2b).  We found
 that for dilution disorder ME and SQF paths are more similar,
  but for a continuous distribution they are usually different, with
  the SQF surfaces being rougher (This can be seen in Fig. 2b). This
  is due to the distributed ``damage'' generated by the SQF process, with
  this process being more important for continuous disorder than for 
  the dilution case. 
    
 Results of simulations in three dimensions are presented in Fig. 3.
 Fig 3a presents data for the roughness of SQF and ME surfaces in
 $40^3$ \{100\} cubic lattices with dilution disorder. We use
 periodic boundary conditions in one direction, and free ones in the 
 other. Two differences between this data and the behavior in
  two dimensions(Fig. 1b) is evident: Firstly SQF surfaces 
  are {\it rougher} than ME surfaces and;
  Secondly all surfaces are quite {\it flat} for 
  weak disorder.  In previous work\cite{n14} we argued that 
  for ME surfaces, there
  is a {\it transition} between a weak disorder phase(with
  perhaps logarithmic roughness) for 
  $p>p_*\sim 0.89$ and an algebraically rough phase $p<p_*$.
    We have done extensive tests of this hypothesis
  with a parallel version of our 
  optimization algorithm\cite{eira} 
  and typical results are presented in
  the inset to Fig. 3a.  In that figure, we present the
  roughness as a function of sample size (up to $400^3$) at $p=0.90>p_*$.
  At $p=0.90$ we expect a log-log plot of the roughness to level off
  indicating an absence of algebraic scaling of the roughness.  This
  is clearly seen in the inset to Fig. 3a and confirms our hypothesis
  of a transition at $p_* < 0.90$. 
 
 A finite-size-scaling plot of the roughness of SQF 
 and ME surfaces in {\it the strong disorder regime} is presented in
 Fig. 3b. This data again shows that, in three dimensions,
  SQF surfaces  are {\it rougher}
 than ME surfaces.
  The ME data(solid symbols) reaches the asymptotic
 exponent $\zeta_{3d}=0.41\pm 0.02$\cite{n14,n15,n16} with
 the sample sizes that are available. Note
 however that at smaller sizes, the $R=1$ ME data has
 a considerably larger slope in this log-log plot.
 The fracture data at $R=1/2$ shows a very simple
 scaling with exponent $0.40\pm 0.05$, which is
 consistent with the ME value. We find that for intermediate
 disorder(e.g. p=0.80) this behavior is typical of SQF surfaces.
 However at stronger disorder the SQF data has strong size
 effects(see $p=0.7$ -  open triangles, and $R=1$ - open squares).
 Although the slope in this data is initially large, 
 it continuously decreases.  We have analysed
 this data in several ways.  Firstly it is
 evident that unless the data changes its
 trend $\zeta < 0.50$.  A more detailed analysis using
 finite size scaling forms, and an analysis of a running
 exponent yields estimates close
 to $0.40$. Though much larger sample sizes are necessary to reach the 
 asymptotic regime for the $R=1$ and $p=0.70$ cases, the data in Fig. 3b
  is consistent with the
 simple conclusion that SQF and ME surfaces have the
 random bond Ising exponents, and that there are
 stronger finite-size effects at stronger disorder.
 
 Now we discuss some reasons for the trends seen in the data of
 Figs. 1-3.  Due to local current concentrations one might expect
 cracks to become ``flatter'' as they become larger. However a random
 void displaced a small vertical amount from a horizontal crack always
 makes the crack deviate, no matter how long the crack (provided the
 neck between the void and the crack is small enough.  Or on lattices
 provided the void is big enough).  Once the
 crack has wandered off the horizontal plane, it has a relatively weak
 memory for the horizontal plane(the stress or current field has
 a rather small gradient on the length scale of the roughness of
 the crack).  This allows the crack to explore the energetically
 most favorable bonds to break, in a similar manner to a ME surface. 
 In fact, there are  mechanisms which can make a SQF surface {\it rougher}
 than a ME one.  In particular, although
 the stress field has a weak {\it average} gradient in the process zone, its
 absolute value is high.  This produces  
 bond breaking in the process zone {\it ahead of the crack tip}.  Since
 the crack propagates through this zone, the disorder it sees is
 larger than that of the pristine disordered system.
  This effect 
 of {\it damage generation} is similar to what is observed in measurements 
 of acoustic emission in slow fracture of 2D media \cite{Yamauchi}.
 Due to this effect SQF surfaces can be  
 rougher than ME surfaces with the same initial disorder.
 Note that although we might expect this mechanism to decrease
 with increasing sample size due to the size effects in damage,
 we are interested in the damage near the crack tip and the behavior
  of that quantity  with sample size is unknown.
  
In summary, we find that SQF and ME surfaces at
strong disorder  
have out of plane roughness exponents
$\zeta_{2d} = 2/3$ and $\zeta_{3d} = 0.41\pm 0.02$.
However at smaller sample sizes our SQF surfaces have 
effective exponent which can be considerabley larger than $0.41$.
  We confirmed that for ME surfaces in 3-d there 
is a weakly-rough to algrebraically-rough 
transition at $p\sim 0.89$.  This implies that
 cracks can be quite flat along the low energy(cusp) 
directions of a crystal lattice.

We thank Edinburgh Parallel Computing Centre
and Center for Scientific Computing, Otaniemi, Finland
for computing resources. This work is 
supported by the EU TRACS scheme, the Technology Development Centre of
Finland, the Academy of Finland (MATRA program
and MA separately), and by DOE grant DE-FG02-090-ER45418(PMD).

\newpage
\begin{center}
{\Large Figure Captions}
\end{center}
 
\noindent{\bf Figure 1. }
 The roughness $w$ of 
 scalar quasistatic fracture surfaces
(SQF) and minimal energy surfaces (ME) 
in the \{10\} orientation of square lattices
with  {\it dilution} (discrete) disorder.
{\bf (a)} Log-log plot of $w$ as a function
 of system size $L$ at $p=0.80$:
SQF (open circles); ME (filled circles).
The dotted line has slope $\zeta = 2/3$.  Boundary conditions
were periodic in the perpendicular direction.  For SQF, 
the number of configurations in the averages, 
$N$, varied from $N=10$ for $L=500$
to $N=256$ for $L=10$, while for the ME  
we used $N=100$ for $L=1000$ up to $N=5000$ for $L=50$.
{\bf (b)} The dependence of $w$ on disorder, 
 for SQF ($L=100, N=30$ - open circles) and ME
  ($L=100, N=5000$ - filled circles). \\

\noindent {\bf Figure 2.} The roughness 
$w$ of scalar quasistatic fracture surfaces
(SQF) and minimal energy surfaces (ME) 
in the  \{10\} orientation square lattices
for a {\it uniform distribution of disorder}.
{\bf (a)} Log-log plot of $w$ versus $L$ for $R=1$:
SQF(open circles); ME (filled circles).
The dotted line has slope $\zeta = 2/3$.  For SQF 
the number of configurations in the averages, 
$N$, varied from $N=50$ for $L=100 $
to $N=400$ for $L=5$, while for the ME case 
we used $N=100$ for $L=1000$ up to $N=5000$ for $L=10$.
{\bf (b)} SQF (diamonds) and ME (plusses) surfaces
 in one configuration of a two-dimensional
\{10\}, $L=100$ random network with $R=1$.\\  

\noindent {\bf Figure 3.}  The roughness, $w$, of SQF and ME surfaces
in three dimensions. {\bf (a)}  $w$ as a function of $p$ for SQF 
($L=40, N=30$ - open circles) and 
ME ($L=40, N=5000$ - solid circles) surfaces.  The inset
shows a finite-size-scaling plot of the roughness at $p=0.90$, which
is in the ``weak disorder'' regime. {\bf (b)}  Log-Log
plot of $w$ in the ``strong disorder'' regime.  Open symbols are for
SQF surfaces and filled symbols are for ME surfaces.  The data
are for: $R=1$(squares); $R=1/2$(circles); $p=0.7$(triangles);
$p=0.50$(diamonds).  As in Figs. 1 and 2 the ME surfaces
are averaged over thousands of configurations, while the largest
size SQF surfaces are averages over around 50 configurations.

   
 
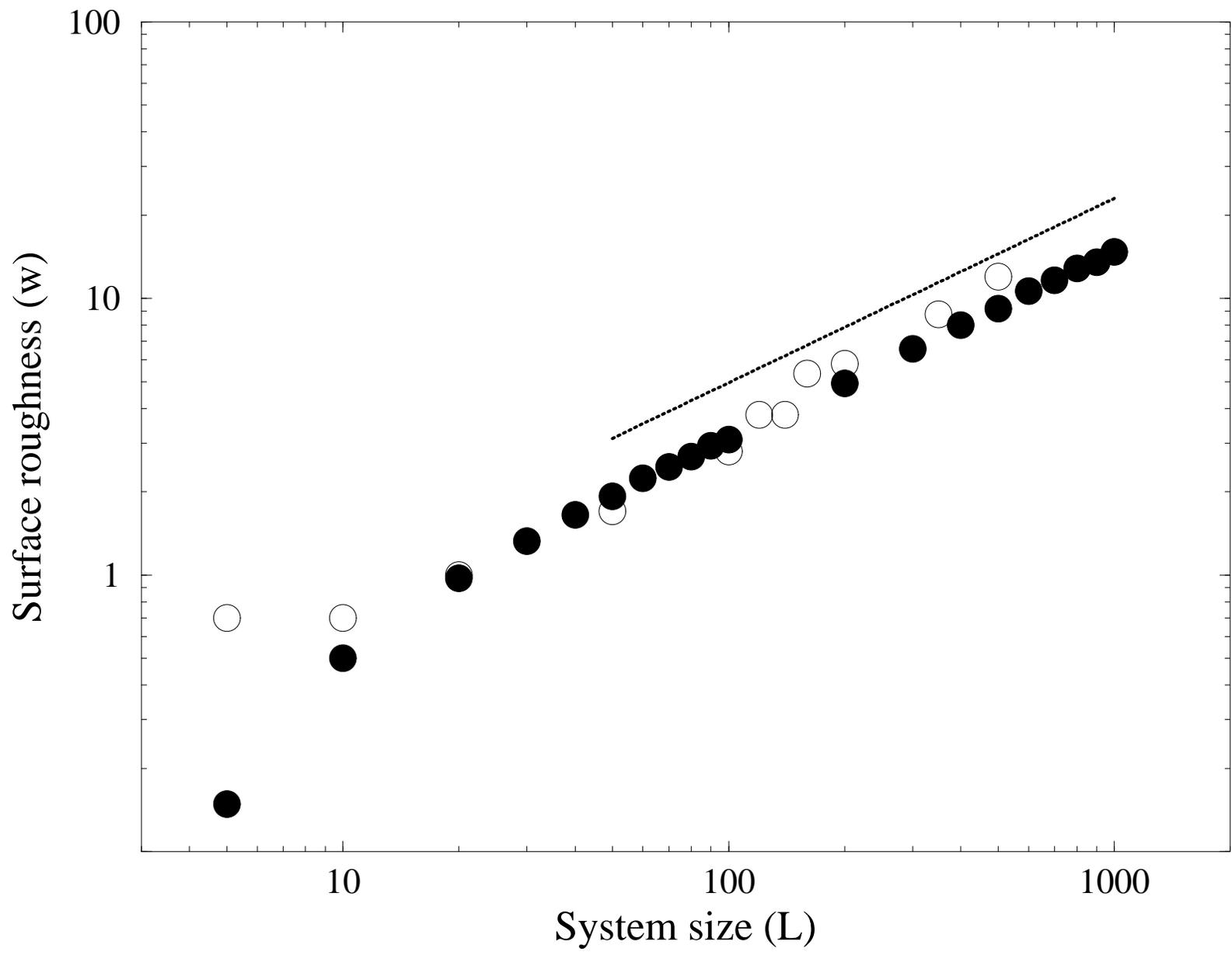
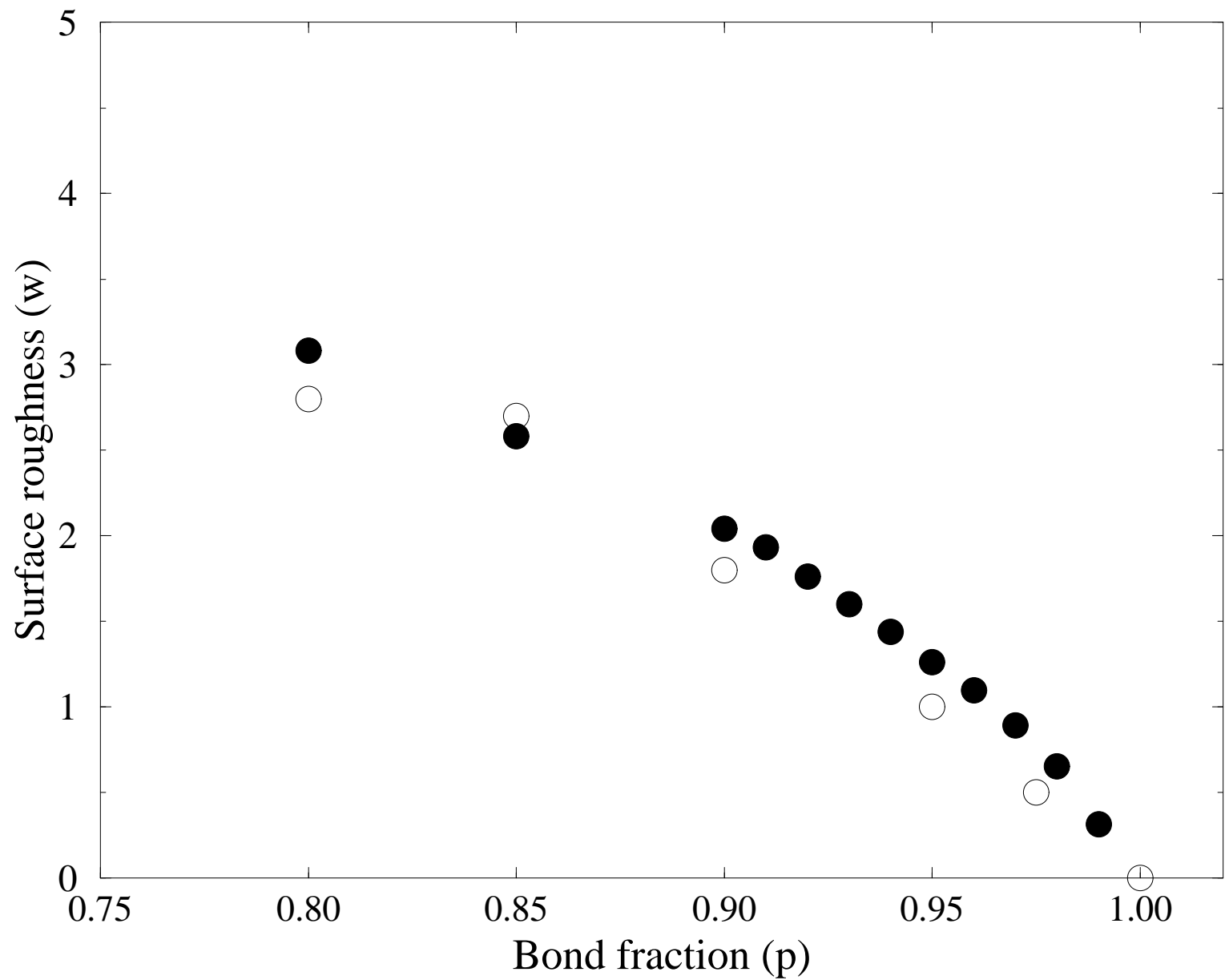
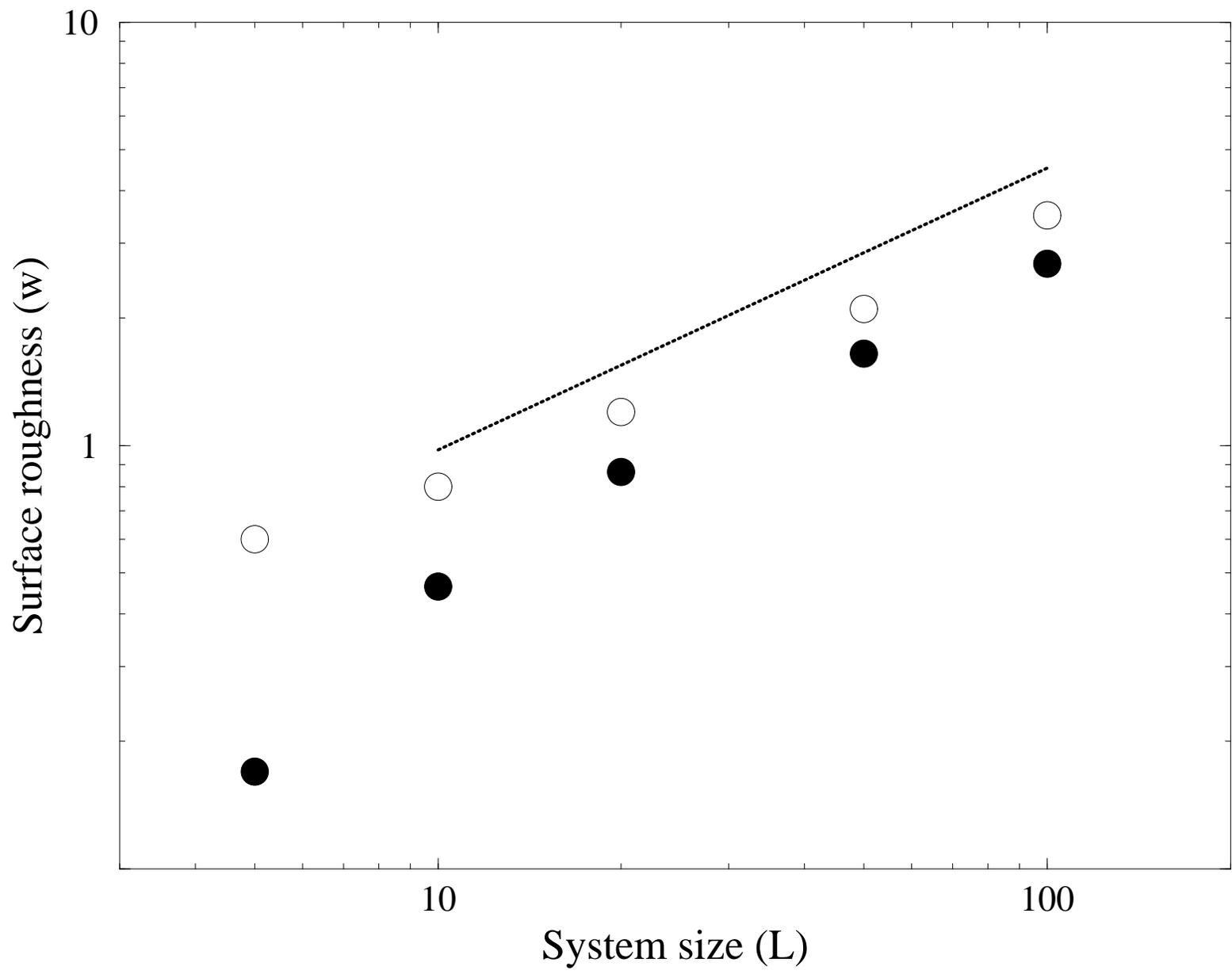
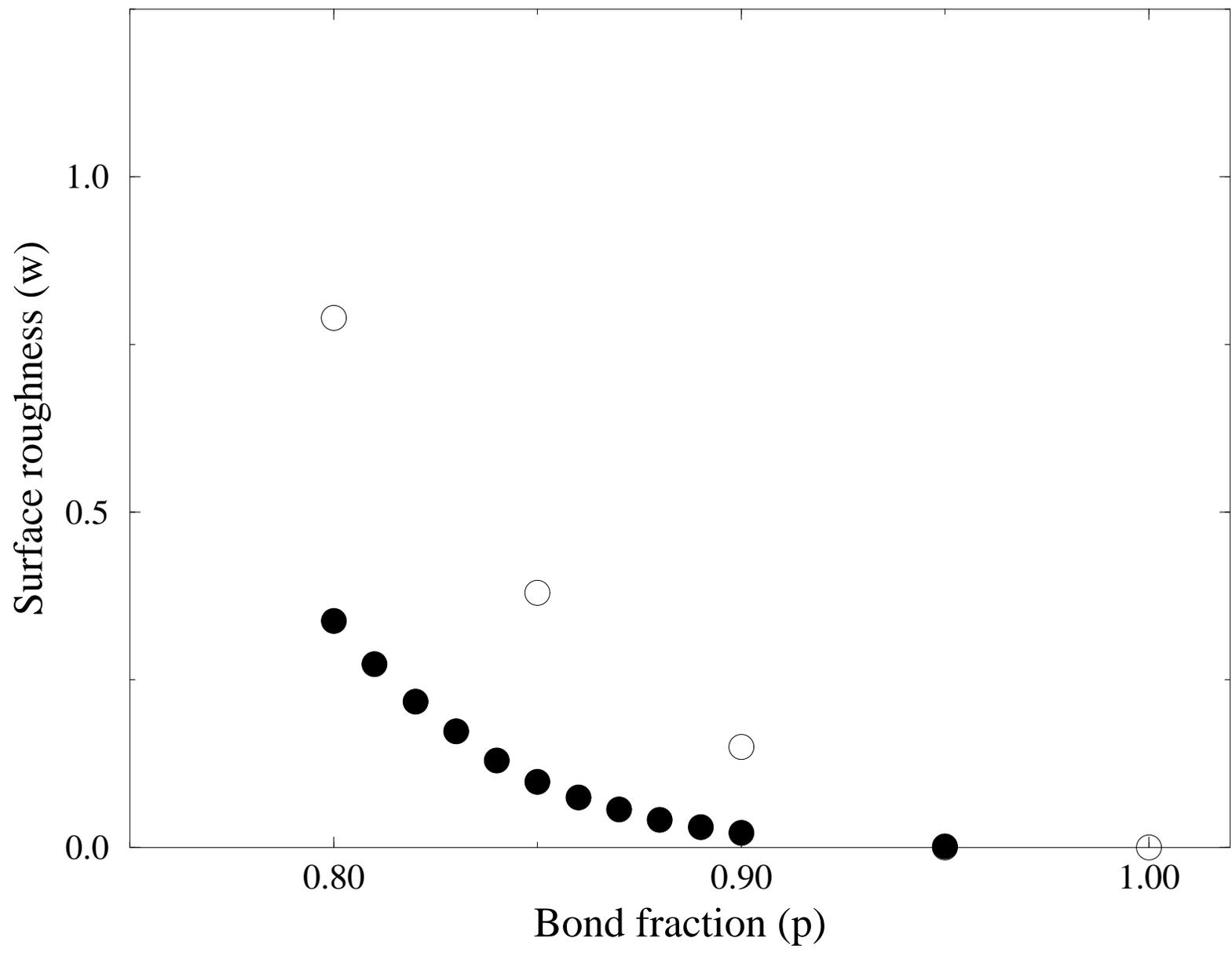
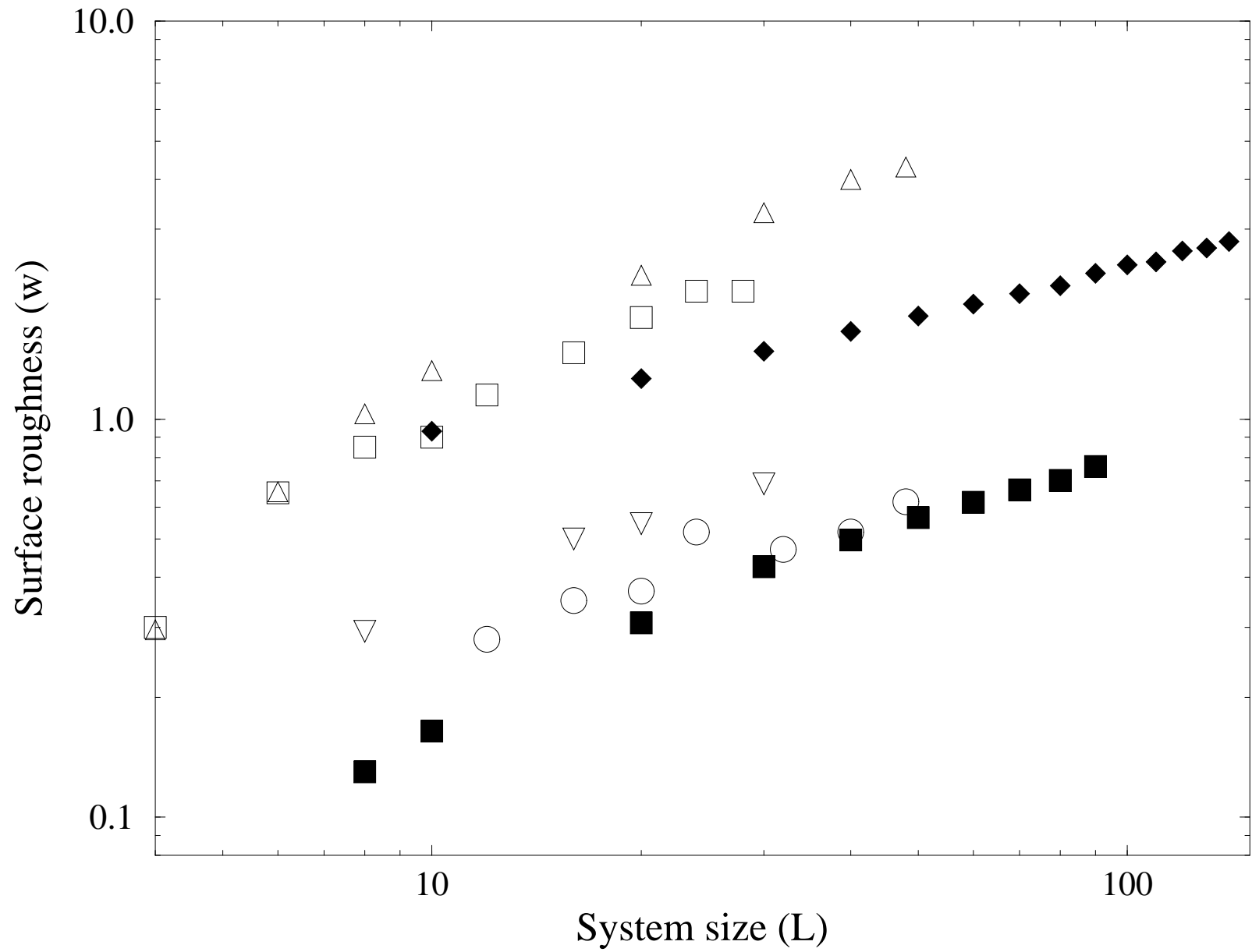
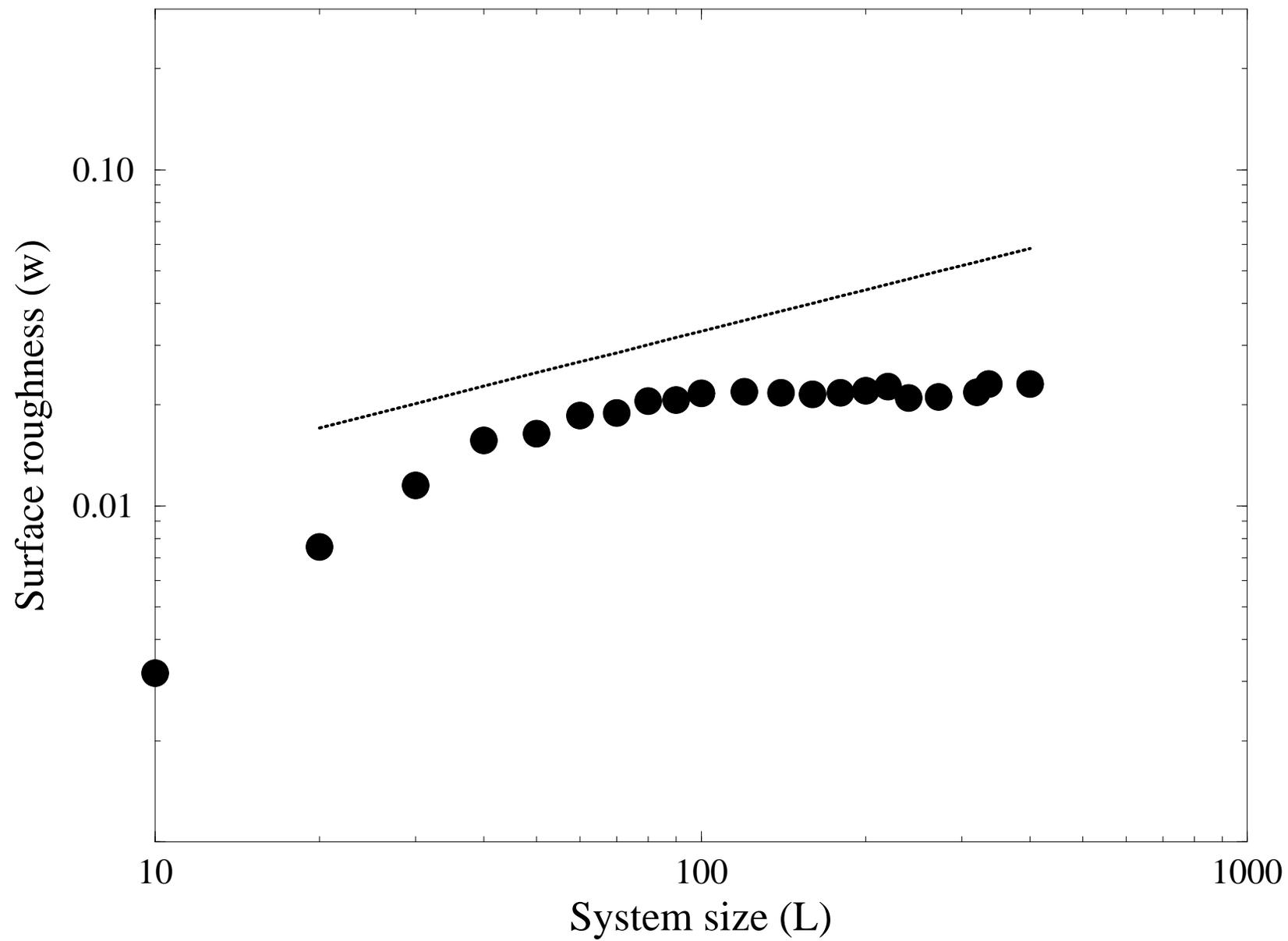
\end{document}